\documentclass[twocolumn,superscriptaddress,prl]{revtex4}
\usepackage{amsfonts,amsmath,amssymb}
\usepackage{graphicx}
\begin{document}
\title{Empirical basis for car-following theory development}
\author{Peter Wagner}
\affiliation{Institute of Transport Research, German Aerospace Center (DLR),
Rutherfordstrasse 2, 12489 Berlin, Germany.}
\author{Ihor Lubashevsky}
\affiliation{Institute of Transport Research, German Aerospace Center (DLR),
Rutherfordstrasse 2, 12489 Berlin, Germany.}
\affiliation{Theory Department, General Physics Institute, Russian Academy of
Sciences, Vavilov Str. 38, Moscow, 119991 Russia}
\date{\today }
\begin{abstract} 
  By analyzing data from a car-following experiment, it is shown that
  drivers control their car by a simple scheme. The acceleration
  $a(t)$ is held approximately constant for a certain time interval,
  followed by a jump to a new acceleration. These jumps seem to
  include a deterministic and a random component; the time $T$ between
  subsequent jumps is random, too. This leads to a dynamic, that never
  reaches a fixed-point ($a(t) \to 0$ and velocity difference to the
  car in front $\Delta v \to 0$) of the car-following dynamics. The
  existence of such a fixed-point is predicted by most of the existing
  car-following theories. Nevertheless, the phase-space distribution
  is clustered strongly at $\Delta v=0$. Here, the probability
  distribution in $\Delta v$ is (for small and medium distances
  $\Delta x$ between the cars) described by $p(\Delta v) \propto
  \exp(-|\Delta v|/\Delta v_0)$ indicating a dynamic that attracts
  cars to the region with small speed differences. The corresponding
  distances $\Delta x$ between the cars vary strongly. This variation
  might be a possible reason for the much-discussed widely scattered
  states found in highway traffic.
\end{abstract} 
 
\maketitle 
 
\section{Introduction} 
 
To understand traffic flow, it is mandatory to analyze the interaction
between the cars. The simplest of these interactions is car-following,
where a car with speed $v$ follows a lead car with speed $V$ in a
certain distance $\Delta x$. Although a lot of theoretical work has
been undertaken to describe this process
(see~\cite{Chowdhury,Helbing,OR-review} for reviews), these models
have a weak empirical basis. Often, calibration and validation is done
only qualitatively, or by comparing the model's output with
macroscopic data like counts and average speeds from induction loop
detectors.
 
The models introduced so far may be classified according to the update
scheme of the underlying dynamic as cellular automata, differential
equations and their time-discretization, and as so called action-point
models \cite{Todosiev1963,Michaels1963}. All models describe the
car-following process by an equation for the acceleration $a$ of the
following car, $a=a\left(\Delta x,v,V \right)$. In case of a
stochastic model, the (white) noise is added to the acceleration,
making $a(t)$ discontinuous in time. Only the action-point models
differ from that, they claim the acceleration more or less constant
until a certain action-point is reached where the acceleration changes
within a short amount of time.
 
In the following, a new data-set will be analyzed that allows a 
detailed analysis of the car-following process. This gives important 
hints how the car-following description can be improved at least on 
the microscopic level.  
 
\section{Results of data analysis} 
 
The data to be analyzed have been described in~\cite{DGPS-RTK}: a
platoon of ten cars, each equipped with differential GPS, drove along
a test track and recorded some hours of car-following data. The time
resolution of the data is 0.1 s, while the accuracy in position of 0.1
m. For each car $n$, its position $x_n(t)$ and speed $v_n(t)$ along
the test track is recorded. The driver of the lead car was instructed
to drive certain speed patterns to estimate the reaction of the
following cars. All drivers were fully informed about the experiment.

Note, that the data contain only the gross headway $\Delta x_n =
x_{n-1} - x_n$, i.e.\ the distance from GPS-receiver to GPS-receiver,
but not the car lengths. From the speed data, the acceleration has
been computed by applying a symmetric Savitzky-Golay
filter~\cite{Press_NR}.
 
\subsection{Distribution of speed differences} 
 
\begin{figure}[ht] 
\begin{center} 
\includegraphics[width=\linewidth]{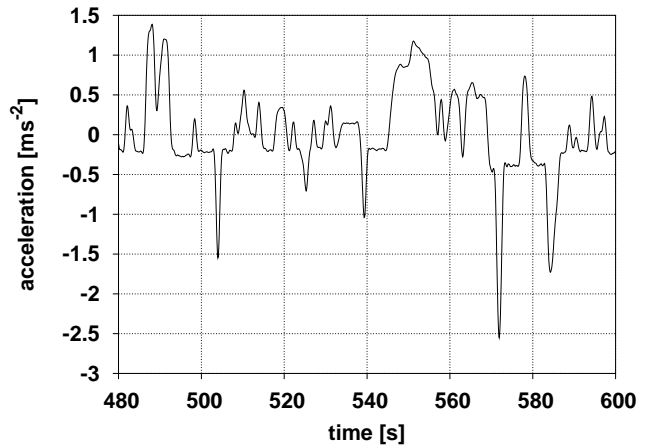} 
\caption{The acceleration $a(t)$ drawn as a function of time. The
  action-points have been determined from $a(t)$ by searching local
  minima/maxima of $\dot a$ where $|\dot a|>\sigma$, where $\sigma$ is
  the noise in $\dot a$. Additionally, to any sequence of points where
  $|\dot a|>\sigma$ only one minimum/maximum has been assigned to
  suppress spurious minima/maxima.}
\label{fig:acces} 
\end{center} 
\end{figure}

The trajectories of the cars in the phase-space $(\Delta v,\Delta x)$
display an oscillatory behavior, see
\cite{Todosiev1963,Brackstone2002} (and references therein) for
similar observations. 
\begin{figure}[h] 
\begin{center} 
\includegraphics[width=\linewidth]{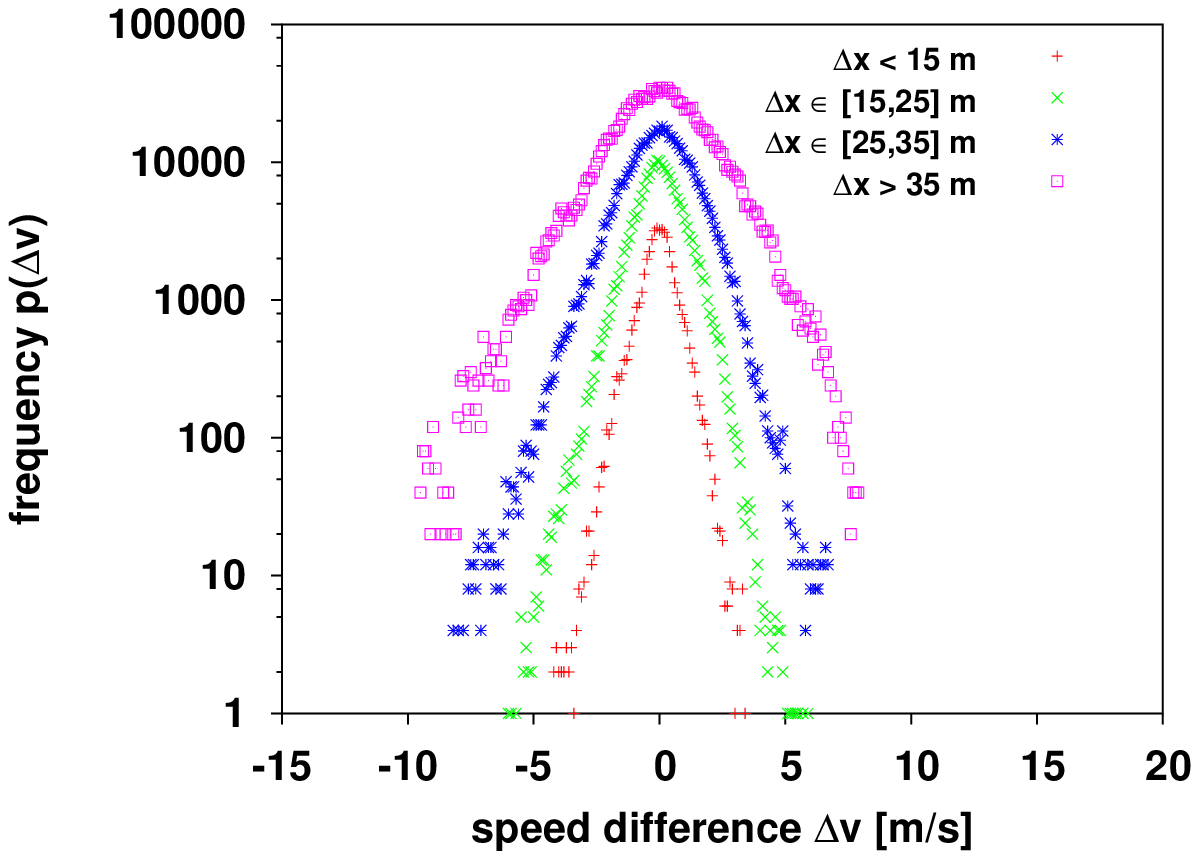}\\ 
\includegraphics[width=\linewidth]{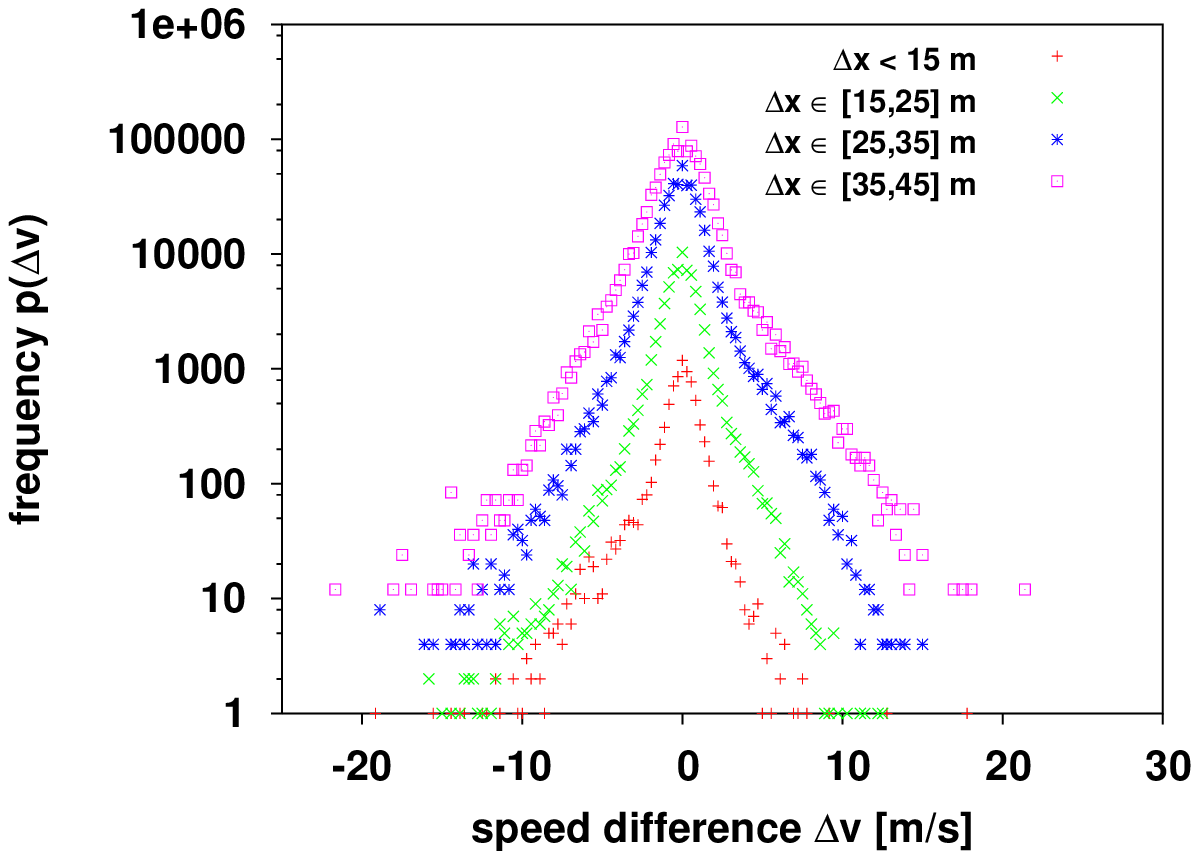} 
\caption{The histogram $p_{\Delta x}(\Delta v)$ of the velocity difference
  for different distances $\Delta x$ between the cars. The upper plot
  is for the GPS data, the lower plot shows the result for single car
  data from the German freeway A1~\cite{Schadschneider-data}. The
  different curves have been scaled to increase visibility.}
\label{fig:pdv} 
\end{center} 
\end{figure} 
However, these oscillations cannot be understood as a noisy
limit-cycle, as can be seen by analyzing the frequency distribution of
the velocity difference $p_{\Delta x}(\Delta v)$. This distribution is
non-analytic for small distances $\Delta x < 30$ m (for the test track
data, for data from the highway this behavior can be observed for even
bigger distances), while there is a cross-over to a more Gaussian
behavior for larger distances, see Fig.~\ref{fig:pdv}. Most
car-following models claim that there is a fixed-point $g^\star$ of
the car-following dynamics where $a=0, \Delta v=0$ and the distance to
the car ahead is given by some function $g^\star(v)$. This idea can be
tested easily with the data at hand, and it is wrong. Although the
cars ``like'' $\Delta v = 0$, they don't get stuck there. The
data-points where $\Delta v$ and $a$ are small (the classical hallmark
of a fixed point) are just part of a very short sequence of one or two
data-points of the time-series of $\Delta v(t)$ and $a(t)$,
respectively. The reason for this can be inferred from the time-series
of the acceleration $a(t)$, see Fig.~\ref{fig:acces}.

This time-series can be understood as a sequence of pieces where 
acceleration is essentially constant (or a linear function of time) 
and fast transitions between these pieces. Since this phenomenon has 
been described already in \cite{Todosiev1963,Michaels1963}, who named 
these points ``action-points'', this name will be used in the 
following. 
 
The pieces where acceleration is not constant, but a linear function of
time can be understood as episodes where the driver does not change
the amount of gas or brake. Because of friction, the acceleration
(positive acceleration increases speed and this increases friction,
lowering acceleration) will change then over time. However, not all
possible values of acceleration are equally likely, e.g.\ a small
negative value (increasing as a function of speed $a_0\propto -c_1 v$)
is chosen much more often. It can be speculated, that these values
belong to a state where the driver is doing nothing, neither touching
gas nor brake.
 
\subsection{Waiting times} 
\begin{figure}[ht] 
\begin{center} 
\includegraphics[width=\linewidth]{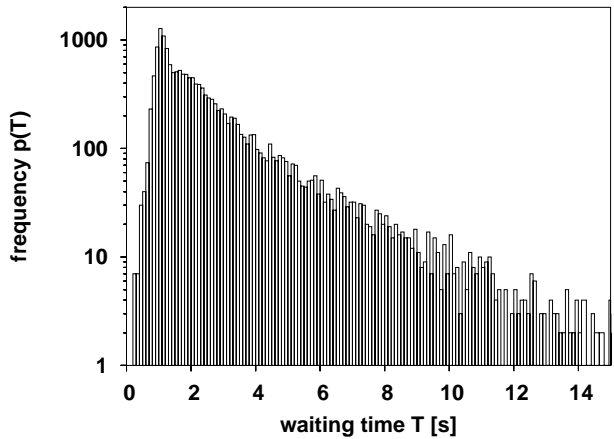} 
\caption{The waiting time distribution between subsequent action-points.} 
\label{fig:prT} 
\end{center} 
\end{figure}

The times $T_i$ between subsequent decisions to change acceleration
have been obtained by the algorithm described in Fig.~\ref{fig:acces}.
In Fig.~\ref{fig:prT}, the corresponding frequency distribution is
displayed.  The exponential tail of this distribution indicates that
any decision is independent from the proceeding one, with one
exception: the probability for making a new decision right after an
action-point is reduced.

More detailed inspections show that the mean-value averaged over all
the drivers is $\langle T \rangle = 2.5\,s$, with large differences
between drivers ranging from $T = 1.8\ldots 4 \,s$. Interestingly, the
waiting time between action-points seems to be independent of distance
or speed difference.
 
\subsection{Acceleration} 
The acceleration data are consistent with a simple linear relationship
\begin{equation} 
a(\Delta x,\Delta v) \propto a_g (g^\star(v) - \Delta x) + a_{\Delta 
v} \Delta v . 
\label{eq:emp-acc} 
\end{equation} 
However, this is only the average behavior, the corresponding standard
deviation $\sigma_a$ of the estimated acceleration values is usually
fairly big. This indicates once more that the whole car-following
process can be described as a stochastic process.
 
\begin{figure}[ht] 
\begin{center} 
\includegraphics[width=\linewidth]{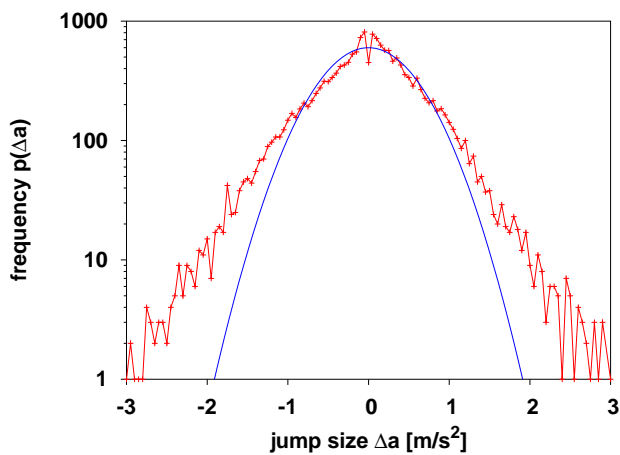} 
\caption{The frequency distribution $p(\Delta a)$ of the acceleration
jumps $\Delta a$. Also shown is a fit of these data to a Gaussian
distribution.}
\label{fig:prob-da} 
\end{center} 
\end{figure}

The analysis of the jumps in $a(t)$ lends further support to this
claim. The jumps are scattered (see Fig.~\ref{fig:action-points}) in
the whole $(\Delta v, \Delta x)$-space. Fig.~\ref{fig:prob-da}
displays the distribution of the jump-sizes. Also in this figure is a
comparison to a Gaussian distribution which indicates that the
jump-size distribution is not Gaussian.

\section{Reflections on modeling} 
 
The results above are inconsistent with most existing microscopic
car-following models. The models claim that the car-following process
tends to converge to some preferred distance $g^\star(v)$ which
depends on speed. In some models, and for some ranges of speed, this
fixed-point is unstable, giving rise to an instability which may
finally be responsible for jam formation. Other models have more than
one fixed point, which has been hypothesized \cite{K2,K2W} as the
reason for the widely scattered states commonly known as synchronized
flow. Since the data above show that there is no such fixed point, on
a microscopic level all these models are inconsistent with the data.
Of course, under a certain amount of course graining a region
$[g^\star_{\text{min}},g^\star_{\text{max}}]$ may serve as a
substitute for a fixed point.  However, definitive conclusions
concerning the macroscopic consequences of the microscopic behavior
found here can be drawn only after suitable models have been
constructed that embed the behavior above in their dynamics.

\begin{figure}[ht] 
\begin{center} 
\includegraphics[width=\linewidth]{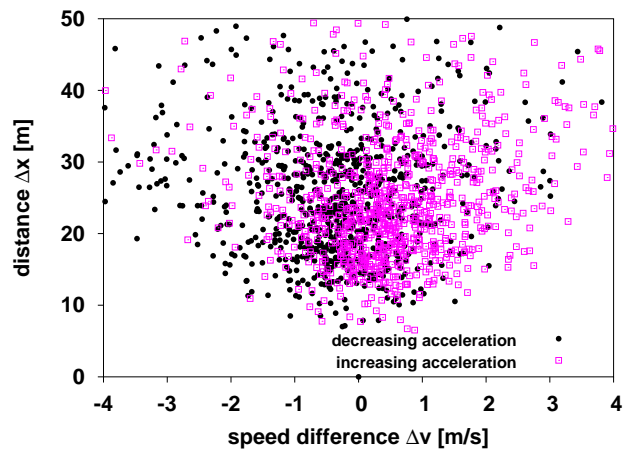} 
\caption{The action-points in the $(\Delta v, \Delta x)$-plane.}  
\label{fig:action-points} 
\end{center} 
\end{figure}

As mentioned already, the so-called action-point models
\cite{Todosiev1963,Michaels1963} behave indeed differently. They are
built around the idea of a constant acceleration between the
action-points, which seems to be in line with the observations above.
Usually, the action-point models include the idea of so-called
perception thresholds in human perception. The most important
threshold $\theta_{\Delta v}$ is the one for determining small changes
in speed difference. It is defined by the equation $\theta_{\Delta v}
\propto \Delta v/(\Delta x)^2$. Although this sounds like a firm
psycho-physical basis, the data above and the data of others
\cite{Brackstone2002} do not support the idea of those thresholds.
Note, however, that the analysis in \cite{Brackstone2002} claims to
have found empirical evidence for those thresholds, however from
inspecting their results different conclusions may be drawn as well.
In Fig.~\ref{fig:action-points}, all the action-points have been drawn
in $(\Delta v,\Delta x)$-space. Their distribution coincides with the
one of all the data-points. Even more, the action-points for
increasing acceleration and the ones for decreasing acceleration cover
almost the same area. Only when action-points with large absolute
values of acceleration are plotted, then some clustering of the points
can be seen.
A more likely interpretation is the idea that the thresholds are
in the space of driving strategies, of which the chosen acceleration
may be a proxy. If the chosen strategy becomes too uncomfortable, then
a new one is chosen. First ideas in this direction have been put
forward already \cite{BRDM,RDM}. 

%There is yet another observation which
%seems to be hard to reconcile at least with the most advanced
%action-point models \cite{Wiedemann1986,Fritz}. There, acceleration is
%assumed to be a deterministic function of the position in state-space
%$\Delta x,v,V$. All the results presented so far in this investigation
%seem to indicate that the car-following process should be understood
%as a stochastic process.
 
From the findings above, more realistic car-following models can be 
constructed. They consist of an acceleration law that determines the 
optimal or desired acceleration as function of speed, speed 
difference, and distance. Based on the analyses performed so far, no 
definitive acceleration model could be favored or ruled out. 
 
So, any of the previously invented car-following models could be used
here, too.  What has to be added, is a specific noise-term that acts
not on acceleration directly, but works on the basis of events: after
a certain randomly chosen time $T$, a new desired acceleration is set.
This desired acceleration has a strong stochastic part, from the data
no unique average (optimal) acceleration could be determined so far.
 
After setting the acceleration to its new value, it remains constant until
the next decision has to be made. The relaxation between two
subsequent values is compatible with a simple relaxation $\dot a =
(a_{\text{des}} - a)/\tau_a$, since the time $\tau_a$ is of the order
of 0.5 s, even the usage of time-discrete models can justified.
Results based on these ideas will be presented in future work.
 
\section{Conclusions} 
 
This work has analyzed experimental car-following data that were
obtained under idealized conditions. Wherever possible, the results
found have been tried to back up by using similar data that have been
obtained from other sources such as highway data. Additional data have
been considered as well, but were not reported here, since they
confirm the analysis done here. Anything taken together, the results
obtained are sufficiently general to support the following
conclusions. The process of car-following can be described by a
relatively simple dynamical process that controls the acceleration.
From time to time, the driver computes a new desired acceleration
(which of course refers to a specific configuration of the gas and
brake pedal) which consists of a deterministic and a random part. The
times between subsequent decisions are randomly distributed, too. The
deterministic part is responsible for the presence of the car near
$\Delta v = 0$ which is a very prominent feature in the data. The
random part competes with this approach to $\Delta v = 0$. It works
that strong, that in fact no fixed point of the car following dynamics
can be established, because acceleration only by chance becomes zero,
and this does not last for a long time.
 
The consequences for microscopic modeling have been derived, 
indications how to construct models that comply with these data have 
been given. A much more difficult task for future research is to find 
out how exactly the choice of strategies is being performed, and how 
big the stochastic component of this choice is.  
 
Finally, the findings presented here may shed a new light on the
discussion about widely scattered states in traffic flow. Simulations
based on a simple model consistent with the findings above display
widely scattered states. We hope to report on this work in future
contributions. However, the question about stability of traffic flow
has to be answered, and it cannot be discussed simply by a linear
stability analysis. The action-point dynamic of human driving on one
hand has a tendency to produce fluctuations that might destroy
homogeneous traffic flow. On the other hand, since the car-following
dynamic converges to a small region in phase-space, at least small
perturbations can heal out. So, the idea that traffic flow is just at
the border of instability got a new meaning.
 
\section*{Acknowledgments} 
 
We are deeply indebted to T.~Nakatsuji and his Hokkaido group for
sharing their data. Those data improved our work and provided valuable
insights because of the ease and elegance to work with them. Other
donations of data came from the Duisburg group of Michael
Schreckenberg, which we acknowledge here as well. These
investigations were supported in part by RFBR Grants~01-01-00389 and
00439. Furthermore, discussions with Boris Kerner, Kai Nagel, and
Andreas Schadschneider helped to clarify the ideas presented here.

\end{document}